\documentstyle[preprint,aps,tighten]{revtex}
\begin{document}
\draft
\title{
Parity Effects and Higher Order Tunneling in
Superconducting SET Transistors
}
\author{Jens Siewert$^1$, Gerd Sch\"on$^1$ and A.D.Zaikin$^{1,2}$}
\address{
$^1$Institut f\"ur Theoretische
 Festk\"orperphysik, Universit\"at Karls\-ru\-he,
 76128 Karlsruhe,  FRG\\
 $^2$I.E. Tamm Department of Theoretical Physics, P.N. Lebedev Physics
Institute, Leninsky Prospect 53, Moscow 117924, Russia
}
\maketitle
\begin{abstract}
Single electron tunneling into small superconducting islands is sensitive to
the gap energy of the excitations created in the process and, hence, depends
on the parity of the electron number in the island.
We study these effects  by analyzing the kinetics of the system.
Since the interplay of Coulomb blockade and parity effects leads to a
blocking of single electron tunneling, higher order tunneling processes
become important. This is well-established for two-electron tunneling.
Here we study processes of third and fourth order. We estimate the
order of magnitude for these processes and discuss their relevance
for recent experiments.
\end{abstract}
\section{Introduction}
Experiments with systems including small superconducting islands
showed effects which depend on the electron number parity in the
island \cite{1,2,3,herg}.
The effect has been interpreted as the even-odd asymmetry predicted by
Averin and Nazarov \cite{AN}. It arises since single electron tunneling from
the ground state, where all electrons near the Fermi surface of the
superconducting island are paired, leads to a state with one extra
electron - the ``odd'' one - in an excited state. In a small island,
where charging effects prevent further tunneling,
the odd electron does not find another excitation for recombination. Hence
the energy of this state stays (at least metastable) above that of the
equivalent normal system by the gap energy. Only at larger gate voltages
another electron can enter the island, and the system can relax to the
 ground state.

At higher temperatures, above a crossover value $T_{cr}$,
 only the $e$-periodic behavior typical for normal metal systems
 has been observed. This crossover has
been explained by free energy arguments in Ref. \cite{1}.
We had developed an alternative approach by studying
the kinetics of the system \cite{szepl}.
Thus,  we can describe nonequilibrium situations
and derive the $I-V$ characteristics of driven systems such as the
SET transistors.
We analyze the rate of tunneling of electrons between the lead and the
island  (reviewed in Section II)  paying particular attention
to the tunneling of the ``odd'' electron (or the tunneling of an
electron into the partner state of the odd electron,
resulting in an instantaneous condensation into a Cooper pair).
The well studied single electron tunneling with rate $\Gamma$
between a superconducting island and lead electrodes
depends on the difference in the energy before and after the
process. This includes the charging energy, but also the excitation
energy $\Delta_i$ in the island. At low
temperature processes which cost energy are suppressed. This renders the
tunneling of the excited ``odd'' electron important.
The tunneling rate $\gamma$ of this single electron is smaller by
a factor $1/N_{eff}$ than the rate $\Gamma$.
Here $N_{eff}$ is the effective number of quasiparticle states available
for an excitation at low temperature;
in mesoscopic islands it is typically of the order of $10^4$.
On the other hand,
in an important range of parameters the rate $\gamma$
is not exponentially
suppressed, since the excitation energy in the island is regained in
this tunneling process. Hence $\gamma \approx \Gamma e^{\Delta_i/k_B T}
/N_{eff}$. Parity effects are observable as long as this single electron
tunneling rate is relevant $\gamma \stackrel{>}{\sim} \Gamma$,
from which we obtain
directly the cross-over temperature
$k_B T_{cr}
\approx \Delta_i/ \ln N_{eff}$.

Our analysis of the tunneling rates can be applied to derive the  $I-V$
characteristics of normal-superconducting-normal (NSN) transistors
(Section III).
In NSN  transistors we combine single electron tunneling
and $2e$-tunneling via Andreev reflection to obtain a richly structured
$I-V$ characteristic, which reproduces quantitatively many features observed
in the experiments of Refs. \cite{3,herg,latest}.
Our $I-V$ characteristic shows wide regions where both single and
two-electron tunneling are blocked due to a combination of the superconducting
gap and the Coulomb blockade,
even at comparatively high transport voltages.
The earlier experiments \cite{herg} showed more  structure, which could be
identified as arising from the influence of the electromagnetic
environment and could be removed by improved shielding \cite{latest}.
Even the latest experiments, however, show more structure than the
theoretical results which account for single and two-electron tunneling.
In particular the experiments show a ridge-like structure with a period $e$ in
the gate charge $Q_g$ at voltages
$eU_{tr} \stackrel{>}{\sim} \Delta_i$.
Cotunneling processes are expected to produce
an unstructured background rather than an effect which
depends in a pronounced way on the gate voltage.
Hence, the question arises whether higher than second order processes
noticeably contribute to the characteristic.
Here we consider coherent three-electron tunneling (Section IV).
The process which turns out to be important is a
coherent combination
of a single electron transition and an Andreev reflection.
The threshold voltage for this process
and the order of magnitude of the rate match well with experimental results.

Another interesting process is cotunneling of two electrons. As this
process is of fourth order in the junction conductance
its contribution is usually small
compared to lower order processes. However, in the limit of small $U_{tr}$
both single electron tunneling and Andreev reflection (as well as their
coherent combination) are suppressed due to
the superconducting gap and Coulomb blockade. In this case cotunneling
processes play the leading role. One of these processes - single electron
cotunneling - has been investigated in Ref. \cite{AN}. The process of
two-electron cotunneling, i.e. the process of two coherent Andreev
reflection events will be discussed in Section V. We will show that in the
limit of low $T$ and $U_{tr}$ either one- or two-electron cotunneling processes
dominate depending on the system parameters. A discussion of our results is
presented in Section VI.

\section{Single electron tunneling rates}
In order to review some basic results we consider an electron box
\cite{2} assuming $\Delta_i < E_C$. The results are easily
extended to cover transport
in SET transistors and systems with $\Delta_i > E_C$. The
electron box is a closed circuit
consisting of a
small superconducting island with total capacitance $C = C_J +C_g$
connected by a tunnel junction with
capacitance $C_J$ to a lead electrode and by a capacitance $C_g$ to a
voltage source $U_g$. The charging energy of the system depends
on $Q_g=C_gU_g$ and the number  $n$ of  charges on the island
\begin{equation}
E_{ch}(n, Q_g) = \frac{(ne-Q_g)^2}{2C} \; .
\label{Ech}
\end{equation}

The normal state conductance of the junction can be
expressed by the tunneling matrix element and the normal density
of states $N_{i/l}(0)$ and volume $V_{i/l}$ of the island and lead:
\[
1/R_t =4\pi e^2 N_i(0)V_iN_l(0) V_l|T|^2/\hbar \ \ \ .
\]
The transition rate $\Gamma^+$ for a single electron tunneling (SET)
process from the lead
to the island, where the number of excess charges on the island changes
from $n$ to $n+1$, is \cite{AL,5}
\begin{equation}
\Gamma^+(Q_g) = \frac{1}{e^2 R_t}
\int_{-\infty}^{\infty} d\xi \int_{-\infty}^{\infty} d\epsilon
{\cal N}_i(\epsilon)
f_l(\xi)[1-f_i(\epsilon)]
\delta(\xi-\epsilon-\delta E_{ch}(Q_g)) \; .
\label{rateint}
\end{equation}
The $\delta$-function expressing energy conservation
depends on the change of the charging
energy before and after the process
 $$
 \delta E_{ch}(Q_g) = E_{ch}(n+1, Q_g) - E_{ch}(n, Q_g) .
 $$

If the distribution functions of lead and island are
equilibrium Fermi functions the expression for the rate reduces to
\begin{equation}
\Gamma^+(Q_g)  = \frac{1}{e} I_{t}(\delta E_{ch}(Q_g))
\frac{1}{\exp[\delta E_{ch}(Q_g)/k_B T] - 1} \; .
\label{rate}
\end{equation}
The SET tunneling rate depends on the difference in the charging energy.
In the superconducting state it depends further on the energy gap,
which enters via the BCS densities of states
${\cal N}_{i}(\epsilon) = \Theta(|\epsilon|-\Delta_{i})
|\epsilon|/\sqrt{\epsilon^2-\Delta_{i}^2}$ of the island
into the well known
quasiparticle tunneling characteristic $I_{t}(eV)$  \cite{Mike}.

As long as the distributions are equilibrium Fermi functions the
rate for the reverse process $\Gamma^-$ is given by the same
expression as (\ref{rate}), however the sign of $\delta E_{ch}$ is
reversed. Both satisfy the condition of detailed balance $\Gamma^-(Q_g) =
\Gamma^+(Q_g) e^{\delta E_{ch}/k_B T}$.
At low temperature in the superconducting state the  rates
$\Gamma^{\pm}$ are large only if the gain in charging
energy exceeds the sum of the energies of the excitation
created in the island  $\pm \delta E_{ch} + \Delta_i
<0 $. They are exponentially suppressed otherwise.

The assumption of equilibrium Fermi distributions is sufficient
as long as we start from the even state.
For definiteness let us assume that we started from $n=0$ and that
the gate voltage
is chosen such that $|Q_g| \le e$. Hence, the rate of tunneling
from an even to an odd state is
\begin{equation}
\Gamma^{eo}(Q_g) = \Gamma^+(Q_g) \; .
\label{Gammaeo}
\end{equation}
However, in a superconductor with an ``odd''
unpaired electron, occupying a quasiparticle state above the gap,
the distribution differs from an equilibrium one. Also
the odd electron can tunnel back to the
lead, enhancing the tunneling from odd to even states. Its initial
energy in the island is at least $\Delta_i$. This makes its tunneling rate
large in a range of gate voltages where the competing processes, the
tunneling of
all the other electrons described by $\Gamma^-$,
are still exponentially suppressed. At finite temperature it is reasonable
to assume that the odd state of the island is described by
a thermal Fermi distribution but with a shifted chemical potential
$f_{\delta \mu}(\epsilon_i) = [e^{(\epsilon-\delta \mu)/T}+1]^{-1}$.
The shift in chemical potential is fixed by the constraint
\begin{equation}
1 = N_i(0) V_i \int_{-\infty}^{\infty} d\epsilon {\cal N}_i(\epsilon)
[f_{\delta \mu}(\epsilon) - f_{0}(\epsilon)] \; .
\label{dmu}
\end{equation}
This reduces at low temperatures to
\begin{equation}
\delta \mu = \Delta_i - T \ln N_{eff},
\label{mu}
\end{equation}
where
 \begin{equation}
 N_{eff}(T)= N_i(0) V_i\sqrt{2\pi\Delta_iT}
\label{neff}
\end{equation}
is the number of states available for quasiparticles near the gap \cite{1}.
The tunneling rate from the odd state to the even state $\Gamma^{oe}$
is given by the expression (\ref{rateint}), however the island
distribution function is replaced by
$f_{\delta \mu}(\epsilon)$. For the following discussion it is useful to
decompose this rate  as
\begin{equation}
\Gamma^{oe}(Q_g) = \Gamma^-(Q_g) + \gamma(Q_g) \; ,
\label{Gammaoe}
\end{equation}
where $\Gamma^-$ has been defined above, and
\begin{equation}
\gamma(Q_g) = \frac{1}{e^2 R_t}\int_{-\infty}^\infty d\epsilon_i
\int_{-\infty}^\infty
d\xi_l {\cal N}_i(\epsilon_i)
[f_{\delta \mu}(\epsilon_i) - f_{0}(\epsilon_i)][1-f_{0}(\xi_l)]
 \delta(\xi_l-\epsilon_i - \delta E_{ch}) \; .
\label{oddrate}
\end{equation}
In the important range of parameters the rate $\gamma$ reduces to
\begin{equation}
\gamma(Q_g) =
\frac{1}{2e^2R_tN_i(0)V_i}
\left\{ \begin{array}{ll}
1
& \mbox{ for\ }  \Delta_i +\delta E_{ch} > T\\
e^{\frac{\Delta_i + \delta E_{ch}}{T}}
& \mbox{ for\ } - \delta E_{ch} -\Delta_i >T
\end{array}
\right.
\label{deltaGamma}
\end{equation}
In comparison to $\Gamma^-$ the odd electron tunneling rate
$\gamma$ contains a small prefactor $1/N_{eff}(T)$.
On the other hand, there exists a range of gate voltages at low
temperature where
$\gamma$ is larger than $\Gamma^-$, since it is not exponentially
suppressed. The
reason is that the gap energy of the island is regained in the process
described by $\gamma$.

For $\exp(-\Delta_i/T) \ll 1$ the ratio of the rates for two
transitions is
\begin{equation}
\Gamma^{oe}/\Gamma^{eo} = \exp[(\delta E_{ch} + \delta \mu)/k_B T] \; .
\end{equation}
The rates $\Gamma^{eo}$
and $\Gamma^{oe} $
obey a detailed balance relation, however, depending on the
free energy difference, which in addition to charging energy
involves the shift of the chemical potential $\delta \mu$ as well.
This free energy difference coincides with that introduced in
Ref. \cite{1}.

Above we described the range
$0 \le Q_g \le e$ where tunneling occurs between the island states $n=0$
and $n=1$. The range $e \le Q_g \le 2e$ can be treated analogously.
The tunneling now connects the states $n=1$ and $n=2$. In this case, except
for the single electron tunneling $\Gamma^{\pm}$, which creates
further excitations,
one electron can tunnel into one specific state ($-k,-\sigma$),
the partner state of the excitation ($k,\sigma$) which
is already present. Both condense immediately; the state with two
excitations only exists virtually. The latter process is described again by
$\gamma(Q_g)$. The symmetry implies
$\Gamma^{eo/oe}(Q_g) = \Gamma^{eo/oe}(2e-Q_g)$.
Since the properties of the system are $2e$-periodic in $Q_g$ we have provided
the description for all values of the gate voltage.

Given the rates $\Gamma^{eo}$ and $\Gamma^{oe}$ we can analyze
where the transition between the even and the odd state occurs.
We can describe the incoherent sequential tunneling
of charges between the island and the lead  by a
master equation for the occupation probabilities of the
even and odd states $W_e(Q_g)$ and $W_o(Q_g)$. It is
\begin{equation}
\frac{d W_e(Q_g)}{dt} = - \Gamma^{eo}(Q_g) W_e(Q_g) +
\Gamma^{oe}(Q_g)W_o(Q_g)
\label{master}
\end{equation}
with $W_e(Q_g)+W_o(Q_g) = 1$. The equilibrium solutions are
\begin{equation}
W_{e(o)}(Q_g) =\Gamma^{oe(eo)}(Q_g)/\Gamma_\Sigma(Q_g),
\label{ws}
\end{equation}
where $\Gamma_\Sigma(Q_g)  =\Gamma^{oe}(Q_g)+\Gamma^{eo}(Q_g)$.
For $\Gamma^{oe} \gg \Gamma^{eo}$ we have $W_e(Q_g) \approx 1$, i.e. the
system occupies the even state, while for $\Gamma^{eo} \gg \Gamma^{oe}$
the island is in the odd state.
Finally, the crossover temperature $T_{cr}$
is determined by $\gamma(e/2) \approx \Gamma[\delta
E^{+}(e/2)])$. The result
\begin{equation}
T_{cr} = \Delta_i/\ln N_{eff}(T_{cr}) \; ,
\label{Tcr}
\end{equation}
 coincides with that found in Refs. \cite{1,2}.
$T_{cr}$ can be interpreted as the temperature, at which the
average number of excitations in the island is at least one.
This result is valid also for SET transistors.

\section{$I-V$ characteristics of SET  transistors}
The analysis presented above can be extended in a straightforward way
to describe even-odd effects in SET transistors, which consist
of an island coupled to two tunnel junctions (left and right)
and a gate capacitance.
In this system the charging energy depends on the gate voltage $U_g$ and
the transport voltage.
The total capacitance of the island $C = C_g + C_l + C_r$
defines the energy scale  $E_C = e^2/2C$. The energy
differences for tunneling processes onto the island
in the left and right junctions are
\begin{equation}
\delta E_{ch,l/r} = E_C - \frac{e(Q_g \pm Q_{tr}/2)}{C} .
\label{Echs}
\end{equation}
For clarity we assume a symmetric bias $U_l=-U_r=U_{tr}$ and $C_l=C_r$.
We further define $Q_{tr} = C U_{tr}$.

We focus our attention on NSN  transistors with an energy gap which is
larger than the charging energy  $\Delta_i > E_C$.
In these systems the odd states have a large energy.
Hence a mechanism which transfers 2 electrons between the normal
metal and the superconductor becomes important. The Andreev
reflection provides such a mechanism \cite{7}.
The  master equation description can be generalized to include also
this process. In the limit considered the rate for Andreev reflection
is given by the same expression as that for single charge tunneling
(\ref{rate}) with the following modifications \cite{GS}:
(i) the charge transferred in an Andreev reflection
is $2e$, and the charging energy changes accordingly,
(ii) the expression for $I_t(V)$ is linear as for normal state
tunneling, but (iii) the effective conductance is of second order
$G^A = R_K/(4N_{ch}R^2_t)$. Here
$R_K = h/e^2$ is the resistance quantum, and the
number of channels $N_{ch} \approx 10^3$ depends on the correlations of the
electron propagation in the lead and the island \cite{HN,ADZ}. An important
conclusion is that  Andreev reflection is also subject to Coulomb blockade
\cite{GS}. Because of the similarity of the rate to that of single
electron tunneling it is clear that the shape of the $I-V$
characteristic due to Andreev reflection also takes a similar form
as in normal transistors.
At low temperatures a set of parabolic current
peaks is found centered around the degeneracy points
$Q_g = \pm e, \pm 3e, \ldots $  \cite{7}
\begin{equation}
I_0^A(\delta Q_g, U_{tr}) = G^A
\Big(U_{tr} - 4\frac{(\delta Q_g)^2}{U_{tr}C^2}\Big) \ \ .
\label{Hekking}
\end{equation}
Here $\delta Q_g = Q_g-e$ for $Q_g$ close to $e$ and similar at the other
degeneracy points.

Depending on the gate and transport voltage we find different mechanisms to
dominate the $I-V$ characteristic. Upon increasing $U_{tr}$ we enter a regime
where single electron tunneling dominates. This ``poisons'' or blocks
the condition for
Andreev scattering  \cite{7}. Hence after an initial increase with
increasing $U_{tr}$ the current suddenly drops at the threshold
voltage
      \begin{equation}
      e U_{tr} = 2 \left( E_C - \frac{e Q_g}{C} + \Delta_i \right)
      \end{equation}
to the value limited by the odd electron tunneling,
i. e. an ``escape'' current of the order
\begin{equation}
I_{esc} \sim \frac{1}{2eR_tN_i(0) V_i} \ \ \ .
\label{creep}
\end{equation}
For a detailed comparison of our calculations with
 available experimental data from Refs. \cite{latest}
we solved the master equation using the parameters of the experiments.

Fig. 1 shows the $I-V$ characteristic as
a function of both gate and transport voltage. At small transport voltage
we find 2e-periodic peaks due to Andreev reflection; the peaks at larger
transport voltages arise due to subsequent incoherent steps of
single electron tunneling
and Andreev reflection processes (which we will refer to as
AQP cycles \cite{latest}). As can be seen in Fig.1 there are
regions of $Q_g$ where even for transport voltages
$U_{tr} \stackrel{>}{\sim} \Delta_i /e$
both single electron and two-electron tunneling is blocked.
In the experiment, however, a current is found in these regions.
Especially, one finds ridge-like structures at
$Q_g = 0, \pm 1, \ldots$ which start at a transport voltage of
$U_{tr} \sim 250 \mu eV$. We argue that these structures
cannot be explained by the presence of several
unpaired quasiparticles or inelastic cotunneling. Consider the
system at $Q_g = 0$. There is no process of first or second order
to alter the number of electrons in the island, hence, it
cannot be explained why
unpaired quasiparticles should appear on the island at the
considered transport voltages. Moreover, increasing the gate voltage
one approaches the
AQP peaks and the probability to
have unpaired quasiparticles increases, so one expects
that a correspondingly large background has to be added to the
current due to AQP processes. The experiment, however,
indicates that instead of such  a background a well-defined
($e$-periodic) structure is added to the features in Fig. 1.
This motivated us to investigate higher order processes.

\section{Coherent three-electron tunneling}
An argument in selecting  the important third-order contributions
is that the processes
may neither produce a large number of excitations nor
change the number of electrons on the island considerably.
This favors processes which are  coherent combinations of
two-electron tunneling in one junction and quasiparticle
tunneling in the other (see the discussion below).

The calculation of the rate is analogous to the calculation
of the two-electron tunneling rate \cite{7}. The rate for a process
with two-electron tunneling in the left junction and one quasiparticle
tunneling in the right can be expressed as
\begin{eqnarray}
\Gamma^{(3)}& =& \frac{2\pi}{\hbar}\sum_{ll'kr}
		 |M_{ll'kr}|^2 f(\xi_l)f(\xi_{l'})(1-f(\epsilon_k))(1-f(\xi_r)
  \times\nonumber
  \\&& \hspace*{2.1cm} \times \delta(\xi_r+\epsilon_k - \xi_l-\xi_{l'}-
                 \frac{3}{2}eU_{tr}+ \delta E_{ch})\ .
\label{rate3}
\end{eqnarray}
It contains the Fermi functions for electrons with energies
$\xi_l$, $\xi_{l'}$
in the left lead, $\xi_r$ in the right lead  and for excitations with
energy $\epsilon_k$ on the island.
We note the important fact that the process has a threshold
voltage
\begin{equation}
eU_{thr} = \frac{2}{3}(\Delta_i +  \delta E_{ch}) \ \ \ .
\label{thresh}
\end{equation}
Below this threshold the rate is exponentially suppressed.
The matrix element $M_{ll'kr}$ has the form
\begin{eqnarray}
M_{ll'kr} & = & \sum_{k'} \frac{T^{(l)}_{l-k'}T^{(l)}_{lk}T^{(r)}_{rk}
			     v_{-k'}u_{k'}v_k}
		   {(\epsilon_k+\epsilon_{k'}+\xi_r-\xi_l-eU_{tr})
		    (\epsilon_k+\xi_r+\delta E_{ch}(-e)-eU_{tr}/2)}
	        \nonumber  \\[2mm]
&&  \hspace{4.7cm}  +\ \  \mbox{5 other terms} \  \  \
\end{eqnarray}
where $u_k, v_k$ denote the coherence factors in the island
and $T^{(l/r)}$ the tunneling matrix elements of the left
and right junctions.
The combination of energy denominators written explicitly
corresponds to quasiparticle tunneling in the first and Andreev
tunneling in the second and third coherent step.
We assume $U_{tr} \approx U_{thr}$ and therefore set
$\xi_l, \xi_{l'}, \xi_r \approx 0$ and $\epsilon_k \approx \Delta_i$.
Since we expect the rate to be important at integer values
of $Q_g/e$, we replace $\delta E_{ch}$ by $E_C$.
Carrying out the integrations and estimating the expressions
for the parameters of the experiment \cite{latest} we find
\begin{equation}
M_{ll'kr} \approx \frac{4\langle T^{(l)}_{l-k'}T^{(l)}_{lk}\rangle_{k'}
				 T^{(r)}_{rk}
		    v_k N_i(0)V_i }{E_C}\  \  \ ,
\end{equation}
where the bracket $\langle \ldots \rangle_p$ denotes averaging over the
directions of $p$.

In the next step, we calculate the voltage dependence
of the rate, assuming that $T$ is much less than any
other energy difference. The result is
\begin{eqnarray}
\Gamma^{(3)} & =& \eta \cdot \left\{\frac{1}{3}
		     \left[ \left( \frac{ \frac{3}{2}eU_{tr}-\delta E_{ch} }
{ \Delta_i } \right)^2  +  2 \right]
                      \sqrt{ \left( \frac{ \frac{3}{2}eU_{tr}-\delta E_{ch} }
		      { \Delta_i } \right)^2  - 1 }  \right.   \nonumber
\\ && \hspace*{1.4cm}  - \left. \left( \frac{ \frac{3}{2}eU_{tr}-\delta E_{ch}
}
       { \Delta_i } \right) \mbox{arcosh}
       \left( \frac{ \frac{3}{2}eU_{tr}-\delta E_{ch} }
	  { \Delta_i } \right) \right\}  \  \ .
\end{eqnarray}
where
\begin{equation}
\eta =   \frac{4\pi}{\hbar}
		      \langle |M_{ll'kr}|^2 \rangle_{ll'kr}
		 N_l(0)^2V_l^2N_r(0)V_rN_i(0)V_i\Delta_i^3
\end{equation}
The momentum average $\langle M_{ll'kr} \rangle_{ll'kr}$
is proportional to the product of the quasiparticle conductance
of the right junction and the Andreev conductance of the left
junction. Using the experimental parameters,
without further assumptions concerning
the averages of the tunneling matrix elements, we estimate $\eta \sim
10^6 \mbox{s}^{-1}$. This  compares well with the
experiments.

The master equation is completed with the rates $\Gamma^{(3)}$
for the various processes which change the island charge by $\pm e$.
The result is shown in Fig. 2. As we have expected there are
no longer regions with blockade of the current at higher
transport voltages. We also note that the ratio
of the heights of Andreev and AQP peaks is changed.
The simulation, however,
does not reproduce the peaked structure of the experimental
results, especially the ridges at $Q_g = 0,\pm 1, \ldots$
are missing.

Finally, we return to the discussion of the criteria to
select the important three-electron contribution. Other
possible realizations are combinations of single electron
tunneling and Andreev reflection {\it in the same} junction,
coherent tunneling of three electrons with the creation of
three quasiparticles
in the island (tunnel events at the same or at different
junctions) and combinations of elastic cotunneling and
single electron tunneling. The first and the second of
these realizations can be excluded, since they
yield much higher threshold voltages than given in (\ref{thresh}).
Furthermore, the experiment shows that elastic cotunneling
plays no role for the $I-V$ characteristic at
transport voltages $eU_{tr} \sim \Delta_i$. This is confirmed
by an estimate using the theory of Ref. \cite{AN}. Hence, higher
order processes including elastic cotunneling can be neglected.

\section{Two-electron cotunneling}

As we have discussed above, the 2$e$-tunneling process
through NS interfaces via Andreev reflection is to a large
extent similar to the process of single electron
tunneling between two normal metals.
We can proceed further with this analogy and
study the process of two-electron cotunneling in NSN transistors
(similarly to the process
of single electron cotunneling in normal tunnel
junctions \cite{AO,AN2,GZ,BOS}).
Let us consider two successive Andreev reflection events at two NS
boundaries of the NSN transistor.
One of these events corresponds to tunneling of two
electrons from a normal metal into
a superconducting island through one of the NS boundaries
(with conversion of two normal electrons into a Cooper pair).
For small values of the external voltage and at low $T$
this process is energetically suppressed because it increases
the charging energy of the
superconducting island by $\sim 4E_C$. The island, however,
can be immediately discharged
due to another coherent tunneling event of a Cooper pair from the island
into the normal metal through the second NS boundary.
Thus Coulomb interaction induces time correlation
between two acts of Andreev reflection at two NS interfaces and the
Coulomb blockade is
lifted for such a process similarly to the case of single electron cotunneling.
For a nonzero transport voltage $U_{tr}$ the above process
provides a finite contribution to the current through the system.
This contribution is proportional to $R_t^{-4}$ but nevertheless
it can become significant in the limit
of small transport and gate voltages, i.e. in the case of Coulomb
blockade of Andreev reflection
at each of the two NS interfaces.

In order to evaluate the two-electron cotunneling current
through NSN transistors
we shall make use of the method developed in Ref. \cite{GZ} for the case of
single electron cotunneling in chains of normal tunnel junctions.
It enables us to evaluate the contribution from
the cotunneling process to the free energy $F_{2e}^{cot}$ and then to calculate
the current by means of the formula
$I_{2e}^{cot}=4e\mbox{Im}F_{2e}^{cot}$.
As the two electron cotunneling mechanism is important
for small transport and gate voltages we restrict
ourselves to the case $Q_g=0$ and $eU_{tr} \ll E_C$. Proceeding
perturbatively  in $\alpha_{t1,2}^A=R_qG^A_{1,2}$ ($G^A_{1,2}$ are the Andreev
conductances of two NS junctions) and assuming that $G^A_{1,2}$
do not depend on
the external voltage in the limit of small $U_{tr}$ one can easily show that
$I_{2e}^{cot} \propto U_{tr}^3$ in complete analogy with single
electron cotunneling
through normal metallic grains \cite{AO,AN2,GZ,BOS}. A rigorous
calculation yields \cite{ADZ}
\begin{equation}
	I_{2e}^{cot}
	=\frac{(4e)^4\alpha^A_{t1}\alpha^A_{t2}U_{tr}^3}{3\pi^7 E_C^2}
	\frac{\Delta_i ^4}{(\Delta_i ^2 - E_C^2)^2}
	\left\{\arctan{\sqrt{\frac{\Delta_i -E_C}
	{\Delta_i +E_C}}}\right\}^4
\label{cot1}
\end{equation}
This result is valid
provided the Andreev conductance of each NS interface is Ohmic. As it was
demonstrated in Refs. \cite{HN,ADZ} under certain physical conditions due
to the proximity effect the Andreev conductance of NS interfaces may become
non-Ohmic $G_{1,2}^A \propto U_{tr}^{-1}$ in the limit of small
voltages. In this
case we find $I_{2e}^{cot}=G^A_{cot}U_{tr}$, where \cite{ADZ}
\begin{eqnarray}
G^A_{cot} & = & \frac{1}
{2\pi^3 e^6 R_{t1}^2R_{t2}^2N_i(0)^2{\cal A}_1{\cal A}_2d_{N1}d_{N2}E_C^2}
\times \nonumber \\
&& \hspace*{2cm} \times \frac{\Delta_i ^4}{(\Delta_i ^2 - E_C^2)^2}
\left\{\arctan\sqrt{\frac{\Delta_i -E_C}
{\Delta_i +E_C}}\right\}^4,
\label{cot2}
\end{eqnarray}
i.e. the two-electron cotunneling conductance $G^A_{cot}$ turns out to be
Ohmic. For reasonable
experimental parameters we can estimate $G^A_{cot} \sim 10^{-10}$
$\Omega^{-1}$. Thus
the effect is in the measurable range and can be important in the limit
of small external voltages.

The effect of environment on two-electron cotunneling can be
treated in complete
analogy with that for the case of single electron
cotunneling \cite{GZ,BOS}. In the case
of Ohmic external impedance $R_x$ the above results for the
cotunneling conductance
should be multiplied by a factor $ \sim (2eU_{tr}R_xC)^{4e^2R_x/\pi}$.

Finally let us point out that the effect of two-electron cotunneling also
occurs in SNS structures. The corresponding cotunneling conductance of
such structures is identical to that for NSN systems with equivalent
parameters.

\section{Discussion}

In conclusion, we have developed a theory of parity effects in small
superconducting islands by analyzing the rates of various tunneling
processes. The tunneling of the single odd excitation created in
earlier tunneling processes  is crucial at low temperatures.
By comparing its rate $\gamma$ to the tunneling
rate $\Gamma^{\pm}$ of all the other electrons we explain the transitions
and the crossover conditions in an electron box.
Our approach can easily be generalized to
derive the $I-V$ characteristics of more complicated systems, such as
SET transistors, which allows us to explain numerous  details
observed in recent experiments. The agreement with experiment
can be improved by including coherent third-order processes.

We considered a superconducting island which always remains coupled
to the leads.
In this respect our model differs from that considered in Refs.
\cite{VA,GZ2,AN3}.
These authors study a system with fixed electron number parity,
either in the even or in the odd
state, while single electron tunneling processes which allow for
transition between the
two states are forbidden. This restricts the excitation spectrum,
leading to
unusual thermodynamic properties. If the island remains coupled to the lead
the excited states alternate between even and odd parity, resulting in rather
different thermodynamic properties.

Our analysis yields a richly structured $I-V$ curves for NSN transistors and
allows us to distinguish different charge transfer mechanisms important for
different values of the external voltage. In particular, we found that
in the low temperature limit the following processes are
important for relatively large gate and/or
transport voltages $Q_g$ and $U_{tr}$:

$a)$ Single electron tunneling (influenced by parity effects) \cite{szepl}

$b)$ Two-electron tunneling (Andreev reflection) \cite{7} and

$c)$ Coherent three-electron tunneling.

We have shown that the correspondence between the simulations
and the experimental results is improved if the theoretical model
takes into account three-electron processes. This model, however,
does not reproduce the ridges in the $I-V$ characteristic in
\cite{latest}. On the other hand, we mention
that the results of Ref. \cite{3} also show no signs of these ridges.

For small values of the external voltage and low $T$ the processes $a)-c)$
are suppressed due to the superconducting gap and Coulomb blockade. Under these
conditions the system conductance is determined
by one of the cotunneling processes:

$d)$ Single electron cotunneling \cite{AN} or

$e)$ Two-electron cotunneling \cite{ADZ}.

As the corresponding contributions to the system conductance depend
on different parameters (in the case $d)$, the conductance
is inversely proportional to $R_t^2N_i(0)\Delta_i$ \cite{AN},
whereas the conductance $e)$ is defined by (\ref{cot1}),
(\ref{cot2})) each of them
can dominate depending on particular experimental situation.

\begin{acknowledgements}
We would like to thank D.V.Averin, M. Devoret, D. Esteve, D.S.Golubev,
F.W.J. Hekking, J. Hergenrother, Y. Nazarov, and M. Tinkham
for stimulating discussions.
  This work is part of ``Sonderforschungsbereich 195'' supported by Deutsche
 Forschungsgemeinschaft. We also acknowledge the support by  INTAS through
 the Grant No 93-790 and by a NATO Linkage Grant.
\end{acknowledgements}

\vspace{1cm}
\noindent
{\large \bf Figure Captions: }
\\[.5cm]
Fig. 1:
The current $I(Q_g,U_{tr})$ through a NSN transistor.
Only single and two-electron tunneling has been taken into account.
The parameters are
chosen to coincide with those of Ref. \cite{latest},
Fig. 3a ($E_C=99\mu eV,
\Delta_i=245 \mu eV$).
\\[5mm]
Fig. 2:
The current $I(Q_g,U_{tr})$ through a NSN transistor including
three-electron tunneling. The parameters are
chosen to coincide with those of Ref.
\cite{latest},
Fig. 3a ($E_C=99\mu eV,
\Delta_i=245 \mu eV$); for the prefactor of the three-electron
rate we set $\eta = 2 \cdot 10^6 s^{-1}$.

\end{document}